\documentclass[cameraready]{Interspeech}

\usepackage{amsmath, amsfonts, amssymb} 
\usepackage{booktabs}
\usepackage{multirow}
\usepackage{makecell}

\usepackage{graphicx}
\usepackage{subcaption}
\usepackage{caption}
\usepackage{graphicx}
\usepackage{subcaption}
\usepackage[export]{adjustbox}

\usepackage{xcolor}

\title{TAD: Token-Adaptive Contrastive Decoding with Confidence-Guided Gating for Hallucination Mitigation in Large Audio-Language Models}

\author[affiliation={1}, orcid=0000-0001-8156-3347, equalcontribution]{Heyu}{Chang}
\author[affiliation={1}, orcid=0000-0003-4619-4325, equalcontribution]{Nianwen}{Si}
\author[affiliation={1}, orcid=0000-0001-8852-6311]{Hao}{Zhang}
\author[affiliation={1}, orcid=0000-0002-6842-9722, correspondingauthor]{Wenlin}{Zhang}
\author[affiliation={1}, orcid=0000-0001-9917-7794, correspondingauthor]{Dan}{Qu}

\address{
    $^1$ Information Engineering University, China
}

\email{okaychy@163.com, snw1608@163.com, haozhang012@163.com, zwlin\string_2004@163.com, qudan\string_xd@163.com}


\usepackage{comment}

\begin{document}

\maketitle

\begin{abstract}

Large audio-language models (LALMs) can hallucinate audio objects, answering ``yes'' to absent sound events, thus undermining reliability in audio question answering. We propose Token-Adaptive Decoding (TAD), a training-free strategy for hallucination mitigation that grounds the initial yes/no decision by contrasting logits under real audio with a matched silent reference. TAD introduces a token-adaptive, confidence-guided gate that is decision-critical at the first decoding step and class-conditional on affirmative tokens, using the audio-silent margin to avoid overcorrection when evidence is weak or already sufficient. Experiments on AudioCaps-Hallucination show that, relative to Audio-Aware Decoding (AAD), a contrastive baseline with fixed contrast strength, TAD improves F1 for Qwen2 by 0.059 to 0.117 across Popular, Adversarial, and Random splits, and for Gemma by 0.025 to 0.064, while on Clotho-AQA it raises F1 from 0.810 to 0.816 on Qwen2 and remains comparable to AAD on Gemma.

\end{abstract}

\section{Introduction}


Large audio-language models (LALMs) \cite{ref12,ref13,ref14,ref22,ref23,ref24} augment text-only large language models with neural audio front-ends, enabling unified conversational systems for acoustic captioning \cite{ref18,ref19} and audio question answering (AQA) \cite{ref20}. Despite their strong general capability, LALMs may generate linguistically plausible yet audio-unsupported responses due to strong language priors and spurious co-occurrences in training data. A prominent and practically consequential failure mode is audio object hallucination: when asked binary presence/absence questions such as ``Is there a siren in the audio?'', LALMs often answer yes even when the queried sound is absent, especially for frequent or semantically compatible categories and under adversarial prompts \cite{ref17}. Recent benchmarks~\cite{ref1,ref3,ref4,ref5} show that state-of-the-art LALMs can capture overall audio content yet still fail on discriminative queries, over-aligning to text priors rather than acoustic evidence---mirroring object hallucination in vision--language models~\cite{ref6,ref7,ref8,ref9}.

To mitigate hallucination without retraining large backbones, prior work has explored training-free inference-time interventions \cite{ref7, ref10}. In the audio domain, Audio-Aware Decoding (AAD)~\cite{ref1} contrasts token logits conditioned on real audio with those under a no-audio (silent) baseline, reweighting predictions to emphasize audio-supported outputs. Adaptive Vector Steering (AVS)~\cite{ref2} injects layer-wise steering vectors derived from paired forward passes on real versus silent audio. These methods demonstrate that contrasting “with-audio” and “without-audio” behavior can substantially reduce hallucination while preserving general AQA performance.


However, prior contrastive approaches use a fixed contrast strength across decoding, failing to adapt to varying audio evidence and thereby limiting hallucination suppression. Moreover, binary AQA outcomes are largely determined at the first decoding step, motivating a decision-aware intervention at this critical point.
To address these limitations, we propose \textbf{Token-Adaptive Decoding (TAD)}, a token-level, confidence–guided decoding method built upon AAD (Figure 1).
Unlike prior methods that apply global reweighting across steps, TAD is decision-critical and class-conditional, reducing overcorrection under weak evidence.
TAD contrasts token logits under real audio and a silent reference, projects them onto \textsc{Yes}/\textsc{No} token sets, and aggregates evidence via log-sum-exp pooling. At the first decoding step, it computes the audio-induced margin gain between \textsc{Yes} and \textsc{No}, and penalizes affirmative tokens only when this gain falls below a threshold, yielding a token-level, decision-aware adjustment of the model’s initial decision without any parameter updates.

\begin{figure}[t]
  \centering
  \includegraphics[width=\linewidth]{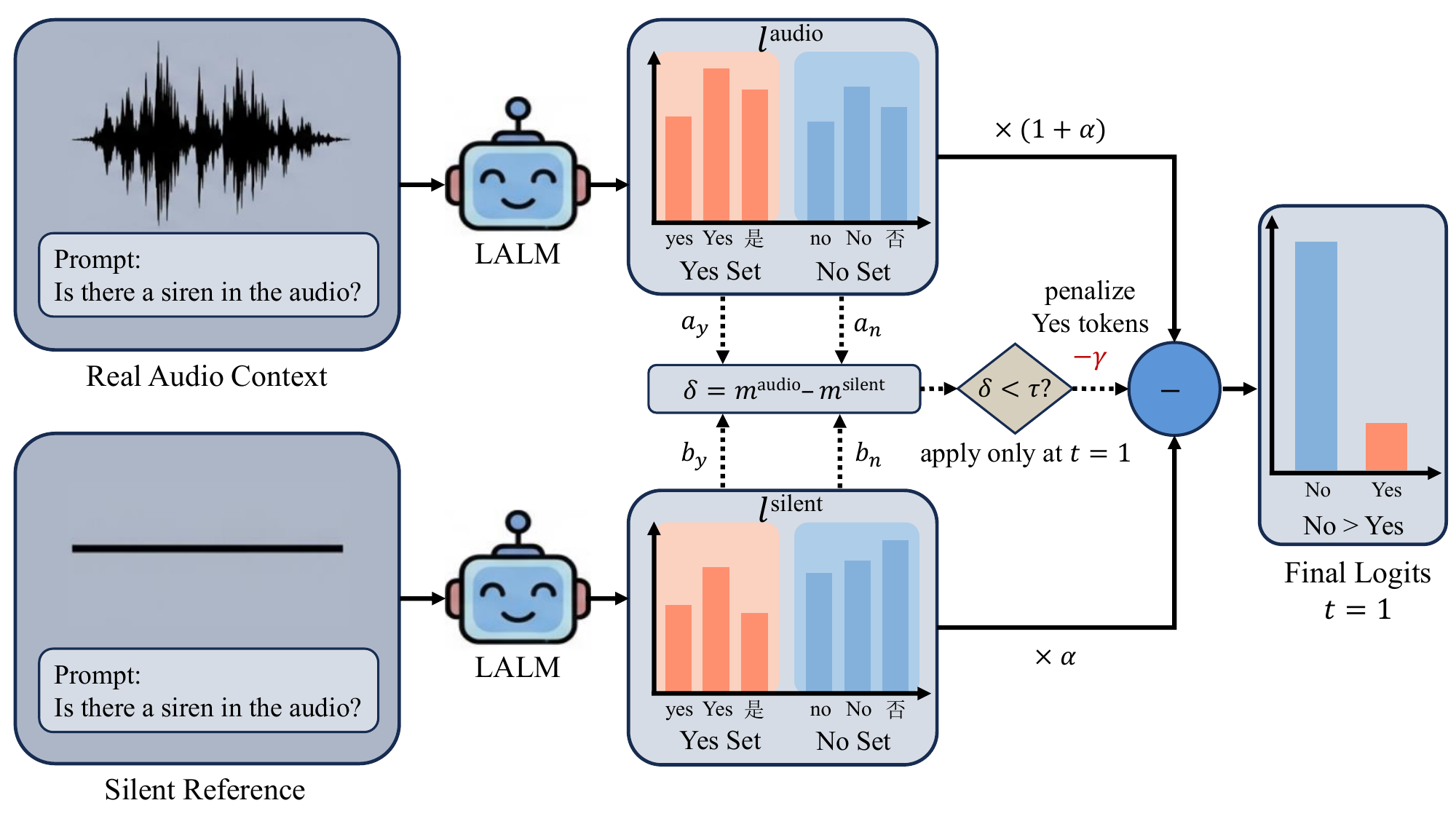}
  \caption{Overview of \textbf{Token-Adaptive Decoding (TAD)} with a silent-reference contrast and a confidence–guided gate for first-step \textsc{Yes}/\textsc{No} decisions.}
  \vspace{-1.5em}
  \label{fig:main}
\end{figure}

Overall, the proposed method provides an audio-driven adjustment of the model's affirmative bias, offering a principled way to suppress audio object hallucination under strong language priors without training. Extensive experiments on hallucination benchmarks (AudioCaps-Hallucination and Clotho-AQA) demonstrate that this simple plug-in logits processor improves hallucination-oriented robustness across diverse settings, while maintaining generally stable AQA performance.

\vspace{-0.3em}
\section{Method}
Given an audio clip $a$ and a natural-language question $x$ about its content, an LALM $f_{\theta}$ generates an answer sequence $\mathbf{y}=(y_1,\ldots,y_T)$ in an autoregressive manner. In audio hallucination settings, we focus on the failure mode where the model produces affirmative responses even when the queried target sound is absent from the audio.

\subsection{Contrastive Decoding with Silent Reference}

We first adopt a decoding strategy based on contrasting predictions under ``with-audio'' and ``without-audio'' conditions. At decoding step $t$, let $\boldsymbol{\ell}^{\text{audio}}_{t}$ denote the logits produced by the model conditioned on the real audio $a$ and the context $(x, \mathbf{y}_{<t})$, and let $\boldsymbol{\ell}^{\text{silent}}_{t}$ denote the logits conditioned on a silent reference $a_0$, constructed as an all-zero waveform with the same length as $a$. The silent branch approximates the model's prediction relying primarily on text and language priors, with minimal effective audio evidence.

Following prior AAD-style formulations~\cite{ref1,ref7,ref21}, we combine the two logits streams to obtain audio-aware contrastive logits:
\begin{equation}
\tilde{\boldsymbol{\ell}}_{t} = (1+\alpha)\,\boldsymbol{\ell}^{\text{audio}}_{t} - \alpha\,\boldsymbol{\ell}^{\text{silent}}_{t},
\end{equation}
where $\alpha \ge 0$ controls the contrast strength. Intuitively, tokens whose scores are substantially higher under real audio than under silence are amplified, while tokens that remain highly scored under silence---indicating strong language-prior influence---are relatively suppressed.

\subsection{Yes/No Token Set Construction}

LALMs output a distribution over the full vocabulary at each step, whereas object hallucination benchmarks typically require a binary decision. We therefore project vocabulary-level logits onto two semantic classes, \textsc{Yes} and \textsc{No}, by constructing token ID sets that cover diverse surface forms of affirmative and negative responses.
However, subword tokenization can map the same semantic word to multiple token fragments. Relying on a single token ID makes the measurement sensitive to tokenization artifacts. To alleviate this, we enumerate a set of surface forms for \textsc{Yes}/\textsc{No}, such as:

\begin{itemize}
    \item \textsc{Yes}: ``yes'', `` yes'', ``Yes'', `` Yes'', etc.;
    \item \textsc{No}: ``no'', `` no'', ``No'', `` No'', etc.
\end{itemize}

For each surface form, we tokenize it without adding special tokens, collect the resulting subword IDs, and take the union with deduplication. This yields the affirmative token set $S_{\text{yes}}$ and the negative token set $S_{\text{no}}$, which are subsequently used to estimate the model's overall tendency toward \textsc{Yes}/\textsc{No} while being robust to specific subword segmentations.

\subsection{Log-Sum-Exp Pooling over Subword Variants}

At each decoding step, we extract aggregated scores for the yes and no semantics from the vocabulary logits
$\boldsymbol{\ell}_t \in \mathbb{R}^{|V|}$.
Given that multiple token IDs may correspond to the same semantic class, we pool logits over
$S_{\text{yes}}$ and $S_{\text{no}}$ using log-sum-exp:
\begin{equation}
\mathrm{pool}(\boldsymbol{\ell}_t, S) = \log \sum_{i \in S} \exp (\boldsymbol{\ell}_{t,i}) ,
\end{equation}
where $S$ is either $S_{\text{yes}}$ or $S_{\text{no}}$.
For numerical stability, we compute
\begin{align}
\mathrm{pool}(\boldsymbol{\ell}_t, S)
&= u + \log \sum_{i \in S} \exp (\boldsymbol{\ell}_{t,i}-u), \\
u &= \max_{i \in S} (\boldsymbol{\ell}_{t,i}) .
\end{align}

Subtracting $u$ prevents numerical overflow in $\exp(\cdot)$ while preserving the exact value of $\mathrm{pool}(\boldsymbol{\ell}_t,S)$, since the constant shift cancels in the log-sum-exp form.

Applying this pooling to both branches yields four scalars:
\begin{align}
a_y &= \mathrm{pool}\big(\boldsymbol{\ell}^{\text{audio}}_{t}, S_{\text{yes}}\big), &
a_n &= \mathrm{pool}\big(\boldsymbol{\ell}^{\text{audio}}_{t}, S_{\text{no}}\big), \\
b_y &= \mathrm{pool}\big(\boldsymbol{\ell}^{\text{silent}}_{t}, S_{\text{yes}}\big), &
b_n &= \mathrm{pool}\big(\boldsymbol{\ell}^{\text{silent}}_{t}, S_{\text{no}}\big).
\end{align}

These pooled scores summarize the model's preference for \textsc{Yes}/\textsc{No} under real-audio versus silent-reference conditions.

\subsection{Confidence-Margin-Guided Gated Decoding}
Building on equation~(1), we introduce a confidence-margin gate applied only at the first decoding step ($t{=}1$), where binary AQA decisions are typically formed. The key idea is to compare the change in the yes-vs-no margin between the audio and silent conditions, and penalize affirmative tokens when audio does not yield a sufficiently positive margin gain.

We first compute the \textsc{Yes}/\textsc{No} margins under real audio and silence:
\begin{equation}
m^{\text{audio}} = a_y - a_n, \qquad
m^{\text{silent}} = b_y - b_n,
\end{equation}
and define the margin gain brought by audio as:
\begin{equation}
\delta = m^{\text{audio}} - m^{\text{silent}}.
\end{equation}

Intuitively, a positive $\delta$ indicates that real audio increases the model's preference for ``yes'' relative to silence, while $\delta \le 0$ suggests that audio provides little evidence supporting an affirmative answer.

We introduce a threshold $\tau$. When $\delta < \tau$, we treat the audio evidence as insufficient and apply an additive penalty $-\gamma$ to all affirmative tokens in the AAD-combined logits $\tilde{\boldsymbol{\ell}}_{t}$:
\begin{equation}
\tilde{\ell}_{t}[i] \leftarrow \tilde{\ell}_{t}[i] - \gamma, \quad \forall i \in S_{\text{yes}},
\end{equation}
where $\gamma > 0$ controls penalty strength. If the condition is not triggered, $\tilde{\boldsymbol{\ell}}_{t}$ remains unchanged. This gating is applied only at $t=1$. Subsequent decoding proceeds solely with the AAD-combined logits.

This design has two key advantages. First, it is \emph{audio-grounded and decision-adaptive}: whether to penalize \textsc{Yes} is determined solely by $\delta$, the audio--silent discrepancy in pooled \textsc{Yes}/\textsc{No} scores, and the penalty is activated only when audio fails to sufficiently increase the \textsc{Yes}--\textsc{No} margin over the silent prior, avoiding fixed-strength overcorrection under weak or already sufficient evidence. Second, it is \emph{task-aligned}: the gate acts only on affirmative tokens, suppressing unsupported \textsc{Yes} outputs while preserving well-formed negative responses.

\section{Experiments}

\subsection{Experimental Setups}
\textbf{Datasets.}
We evaluate TAD on two AQA benchmarks.
\textit{i}) AudioCaps-Hallucination~\cite{ref3} is built from the AudioCaps test split~\cite{ref16} by generating binary questions from five templates. For each template, the queried object is sampled with three strategies (\textsc{Random}, \textsc{Adversarial}, \textsc{Popular}), yielding 30,220/31,047/31,376 QA pairs, respectively, and is designed to test rejection of absent objects.
\textit{ii}) Clotho-AQA~\cite{ref15} is an AQA benchmark of environmental audio clips. In our evaluation, we focus on the subset of 1,991 samples with questions whose answers are ``yes'' or ``no''~\cite{ref1}.

\textbf{Models.}
We evaluate on two LALMs with different model scales:
\textit{i}) Qwen2-Audio-7B-Instruct~\cite{ref12}, an instruction-tuned~\cite{ref25} audio--language model equipped with an audio encoder; and
\textit{ii}) Gemma-3n-E4B-it~\cite{ref13}, a lightweight instruction-tuned~\cite{ref25} multimodal model with $\sim$4B parameters.

\textbf{Metrics and hyperparameters.}
Following prior works~\cite{ref1,ref2}, we treat \textsc{No} as the positive class, corresponding to correctly rejecting absent objects,  and report accuracy, precision, recall, and F1 from final responses. The contrast weight $\alpha$ controls audio--silent reweighting and is typically set to 0.5 or 1.0~\cite{ref1}. We use $\tau{=}0.2$ within $[0,0.5]$ and $\gamma{=}2.5$ within $[2,4]$. We do not exhaustively ablate $\alpha$, $\tau$, and $\gamma$, focusing instead on overall effectiveness and robustness. We use the following prompt: "\textit{Focus on the given audio and answer the following question. Answer in the format: 'Yes, ...' or 'No, ...', and always start with 'Yes' or 'No'.}" Our code is available at: {\url{https://github.com/Changhy26/TAD}}.

\subsection{Main Results}

\subsubsection{Results on AudioCaps-Hallucination}

Tables~1 and~2 report results on AudioCaps-Hallucination under \textsc{Random}, \textsc{Adversarial}, and \textsc{Popular} settings. We primarily compare F1 and recall, since mitigating affirmative hallucinations requires improving the ability to answer ``no'' when the queried event is absent, while keeping accuracy competitive. For Qwen2-Audio-7B-Instruct in Table~1, default decoding exhibits a pronounced affirmative bias, with high precision but low recall and an F1 of only $0.395$ on \textsc{Random}. Both AAD and AVS substantially improve over default decoding. Our TAD achieves the best F1 on \textsc{Random} and \textsc{Popular}, reaching $0.853$ and $0.558$, driven by higher recall, which indicates more reliable rejection of unsupported ``yes'' responses. On \textsc{Popular}, TAD attains the highest recall and F1 while maintaining accuracy on par with competing methods, suggesting stronger robustness to frequent but absent categories. AVS yields the highest F1 on \textsc{Adversarial} at $0.524$, but TAD remains competitive and improves over AAD, indicating that the proposed decision-aware intervention offers gains across sampling regimes rather than relying on a fixed contrast strength.

For Gemma-3n-E4B-it in Table~2, the same methods improve over default decoding but with a clearer conservativeness--accuracy trade-off. AVS performs best on \textsc{Random}, reaching an F1 of $0.727$. TAD yields the highest F1 on the more challenging \textsc{Adversarial} and \textsc{Popular} splits, achieving $0.611$ and $0.582$. Increasing $\alpha$ generally increases recall for TAD on harder settings, but the effect is not uniformly beneficial, as seen from the slight F1 drop on \textsc{Random}, suggesting that overly strong contrast can over-correct when audio evidence is already sufficient.

\begin{table}[t]
\centering
\caption{Results of Qwen2-Audio-7B-Instruct on the AudioCaps-Hallucination dataset.}
\label{tab:qwen2audiocaps_halluc}
\scriptsize
\setlength{\tabcolsep}{4.5pt}      
\begin{tabular}{llccccc}
\toprule
Division & Method & $\alpha$ & Accuracy & Precision & Recall & F1 \\
\midrule
\multirow{6}{*}{Random}
& Default & --  & 0.593 & 0.769 & 0.266 & 0.395 \\
& AVS~\cite{ref2}     & --  & 0.773 & 0.706 & 0.937 & 0.805 \\
& \multirow{2}{*}{AAD~\cite{ref1}} & 0.5 & 0.705 & 0.843 & 0.504 & 0.631 \\
&                     & 1.0 & 0.762 & 0.824 & 0.666 & 0.736 \\
& \multirow{2}{*}{TAD (ours)} & 0.5 & 0.797 & 0.844 & 0.729 & 0.782 \\
&                     & 1.0 & \textbf{0.841} & \textbf{0.847} & \textbf{0.858} & \textbf{0.853} \\
\midrule
\multirow{6}{*}{Adversarial}
& Default & --  & 0.469 & 0.399 & 0.217 & 0.281 \\
& AVS~\cite{ref2}     & --  & \textbf{0.481} & \textbf{0.503} & \textbf{0.548} & \textbf{0.524} \\
& \multirow{2}{*}{AAD~\cite{ref1}} & 0.5 & 0.462 & 0.425 & 0.325 & 0.369 \\
&                     & 1.0 & 0.441 & 0.441 & 0.411 & 0.425 \\
& \multirow{2}{*}{TAD (ours)} & 0.5 & 0.466 & 0.445 & 0.471 & 0.457 \\
&                     & 1.0 & 0.465 & 0.451 & 0.547 & 0.494 \\
\midrule
\multirow{6}{*}{Popular}
& Default & --  & 0.503 & 0.474 & 0.291 & 0.361 \\
& AVS~\cite{ref2}     & --  & \textbf{0.547} & \textbf{0.566} & 0.546 & 0.555 \\
& \multirow{2}{*}{AAD~\cite{ref1}} & 0.5 & 0.499 & 0.483 & 0.408 & 0.442 \\
&                     & 1.0 & 0.483 & 0.493 & 0.506 & 0.499 \\
& \multirow{2}{*}{TAD (ours)} & 0.5 & 0.504 & 0.487 & 0.553 & 0.518 \\
&                     & 1.0 & 0.510 & 0.493 & \textbf{0.643} & \textbf{0.558} \\
\bottomrule
\end{tabular}
\end{table}
\vspace{-0.5em}

\begin{table}[t]
\centering
\caption{Results of Gemma-3n-E4B-it on the AudioCaps-Hallucination dataset.}
\label{tab:gemma3n_audiocaps_halluc}
\scriptsize
\setlength{\tabcolsep}{4.5pt}      
\begin{tabular}{llccccc}
\toprule
Division & Method & $\alpha$ & Accuracy & Precision & Recall & F1 \\
\midrule
\multirow{6}{*}{Random}
& Default & --  & 0.664 & \textbf{0.711} & 0.553 & 0.622 \\
& AVS~\cite{ref2}     & --  & \textbf{0.693} & 0.654 & 0.819 & \textbf{0.727} \\
& \multirow{2}{*}{AAD~\cite{ref1}} & 0.5 & 0.658 & 0.648 & 0.704 & 0.675 \\
&                     & 1.0 & 0.534 & 0.612 & 0.810 & 0.697 \\
& \multirow{2}{*}{TAD (ours)} & 0.5 & 0.638 & 0.586 & 0.939 & 0.722 \\
&                     & 1.0 & 0.589 & 0.572 & \textbf{0.938} & 0.710 \\
\midrule
\multirow{6}{*}{Adversarial}
& Default & --  & 0.471 & 0.434 & 0.348 & 0.386 \\
& AVS~\cite{ref2}     & --  & 0.489 & \textbf{0.508} & 0.631 & 0.563 \\
& \multirow{2}{*}{AAD~\cite{ref1}} & 0.5 & 0.467 & 0.452 & 0.529 & 0.488 \\
&                     & 1.0 & 0.403 & 0.462 & 0.671 & 0.547 \\
& \multirow{2}{*}{TAD (ours)} & 0.5 & \textbf{0.492} & 0.482 & 0.825 & 0.608 \\
&                     & 1.0 & 0.474 & 0.483 & \textbf{0.831} & \textbf{0.611} \\
\midrule
\multirow{6}{*}{Popular}
& Default & --  & 0.483 & 0.455 & 0.377 & 0.412 \\
& AVS~\cite{ref2}     & --  & \textbf{0.515} & \textbf{0.527} & 0.628 & 0.573 \\
& \multirow{2}{*}{AAD~\cite{ref1}} & 0.5 & 0.472 & 0.460 & 0.542 & 0.498 \\
&                     & 1.0 & 0.417 & 0.477 & 0.670 & 0.557 \\
& \multirow{2}{*}{TAD (ours)} & 0.5 & 0.459 & 0.462 & 0.756 & 0.574 \\
&                     & 1.0 & 0.449 & 0.469 & \textbf{0.767} & \textbf{0.582} \\
\bottomrule
\end{tabular}
\end{table}
\vspace{-0.5em}

\subsubsection{Results on Clotho-AQA}

Table~3 reports results on Clotho-AQA. For Qwen2-Audio-7B-Instruct, both AAD and TAD improve over default decoding mainly through higher recall and F1, indicating more reliable rejection of unsupported objects. TAD achieves the best F1 of $0.816$ with $\alpha{=}1.0$ while keeping accuracy competitive, suggesting that the proposed decision-aware first-step gating strengthens audio-grounded decisions without harming overall response quality. For Gemma-3n-E4B-it, AAD and TAD similarly boost recall from $0.470$ to at least $0.844$, but this increased conservativeness comes with reduced precision and accuracy, reflecting a stronger tendency to answer ``no''. Even with this trade-off, both methods deliver clear F1 gains over default decoding, with AAD reaching $0.658$ and TAD remaining close at $0.650$, indicating that the benefits of the intervention persist for a smaller model.

\begin{table}[t]
\centering
\caption{Results of Qwen2-Audio-7B-Instruct and Gemma-3n-E4B-it on the Clotho-AQA dataset.}
\label{tab:qwen2_gemma_audiocaps_halluc}
\scriptsize
\setlength{\tabcolsep}{4.5pt}      
\begin{tabular}{llccccc}
\toprule
Model & Method & $\alpha$ & Accuracy & Precision & Recall & F1 \\
\midrule
\multirow{5}{*}{\makecell[l]{Qwen2-Audio\\-7B-Instruct}}
& Default & --   & 0.759 & \textbf{0.805} & 0.597 & 0.686 \\
& \multirow{2}{*}{AAD~\cite{ref1}}
         & 0.5 & 0.785 & 0.796 & 0.791 & 0.789 \\
&        & 1.0   & 0.812 & 0.755 & 0.873 & 0.810 \\
& \multirow{2}{*}{TAD (ours)}
         & 0.5   & \textbf{0.818} & 0.741 & 0.900 & 0.813 \\
&        & 1.0   & 0.796 & 0.741 & \textbf{0.907} & \textbf{0.816} \\
\midrule
\multirow{5}{*}{\makecell[l]{Gemma-3n\\-E4B-it}}
& Default & --  & \textbf{0.662} & \textbf{0.664} & 0.470 & 0.551 \\
& \multirow{2}{*}{AAD~\cite{ref1}}
         & 0.5 & 0.629 & 0.575 & 0.719 & 0.639 \\
&        & 1.0 & 0.557 & 0.539 & 0.844 & \textbf{0.658} \\
& \multirow{2}{*}{TAD (ours)}
         & 0.5 & 0.600 & 0.527 & 0.847 & 0.650 \\
&        & 1.0 & 0.524 & 0.518 & \textbf{0.860} & 0.647 \\
\bottomrule
\end{tabular}
\end{table}

\subsection{Further Analysis}
\subsubsection{Confusion-matrix Perspective}

\begin{figure}[t]
  \centering

  \begin{subfigure}[t]{0.15\textwidth}
    \centering
    \includegraphics[width=\linewidth]{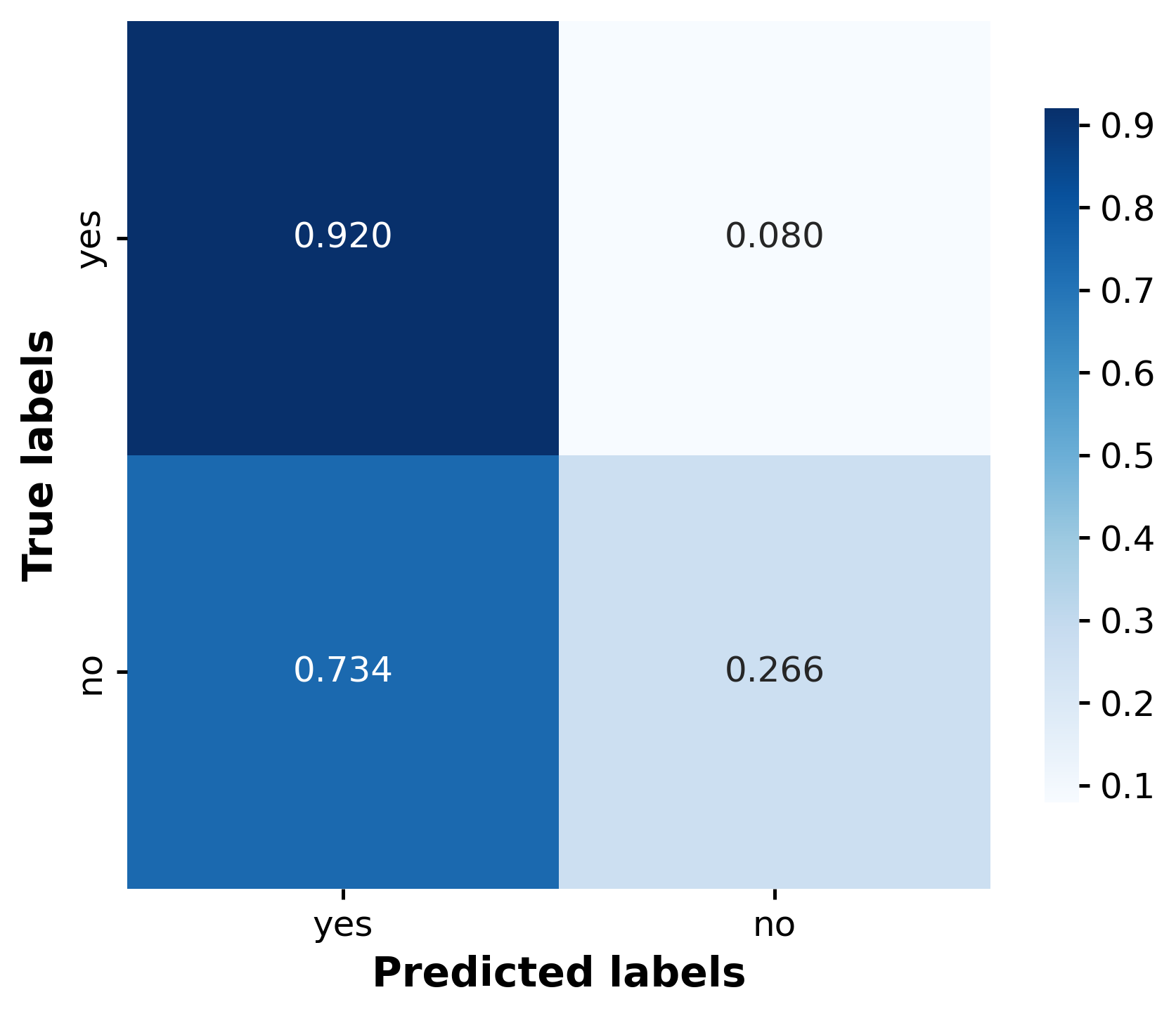}
    \caption{Default}
    \label{fig:fig2_default}
  \end{subfigure}\hfill
  \begin{subfigure}[t]{0.15\textwidth}
    \centering
    \includegraphics[width=\linewidth]{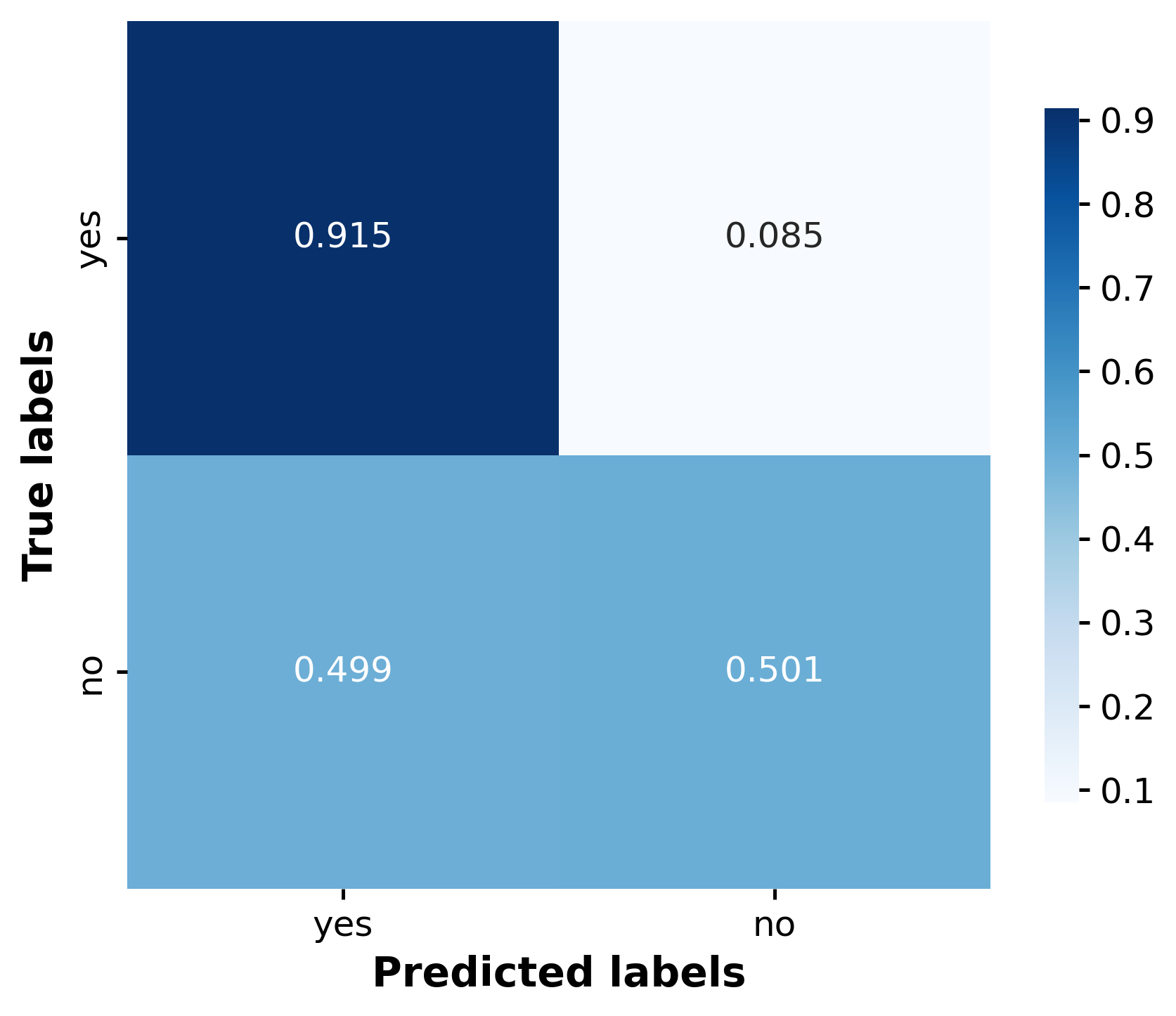}
    \caption{AAD ($\alpha=0.5$)}
    \label{fig:fig2_aad_05}
  \end{subfigure}\hfill
  \begin{subfigure}[t]{0.15\textwidth}
    \centering
    \includegraphics[width=\linewidth]{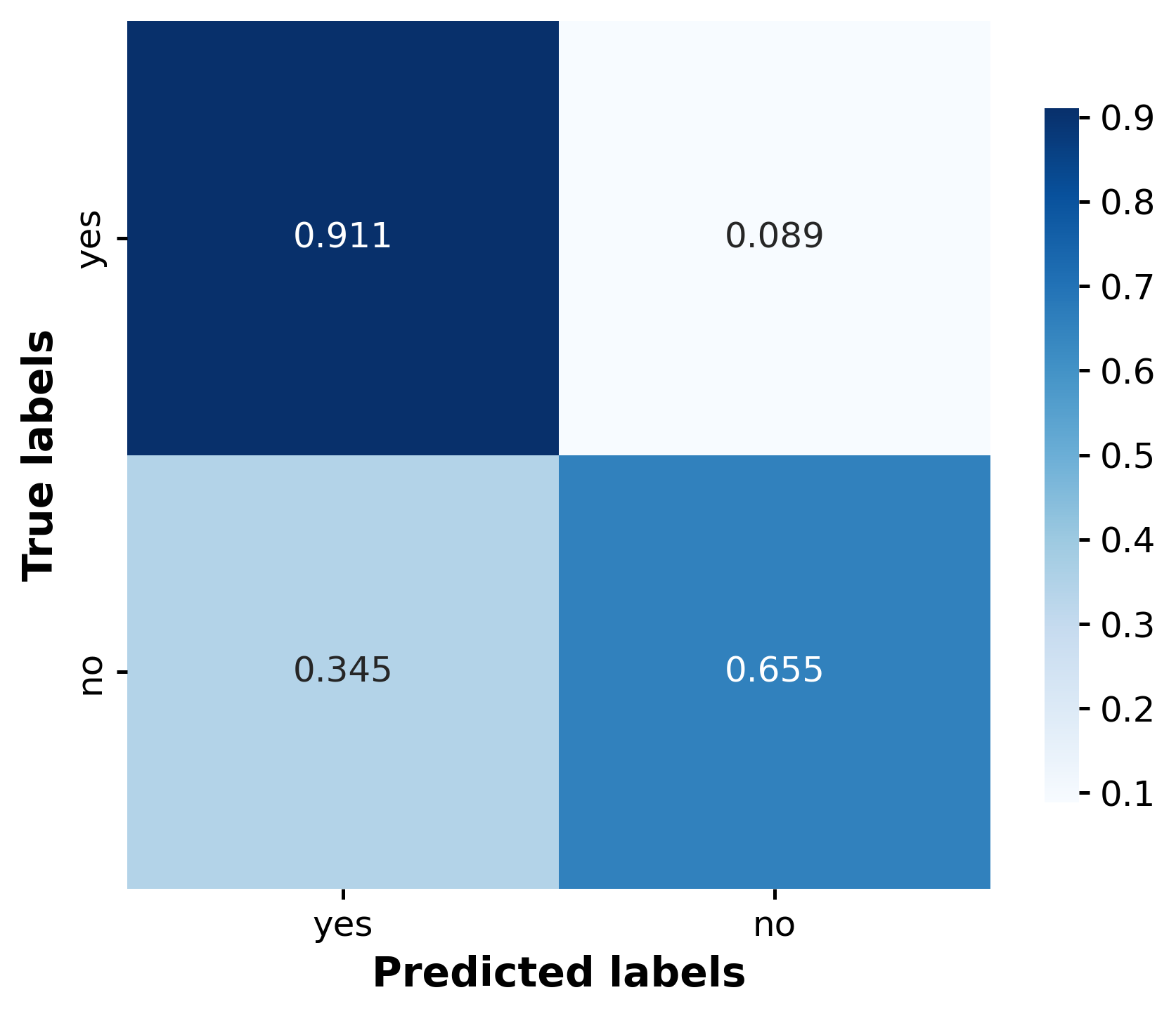}
    \caption{AAD ($\alpha=1.0$)}
    \label{fig:fig2_aad_10}
  \end{subfigure}

  \vspace{0.6em}

    \hspace*{\fill}
    \begin{subfigure}[t]{0.15\textwidth}
      \centering
      \includegraphics[width=\linewidth]{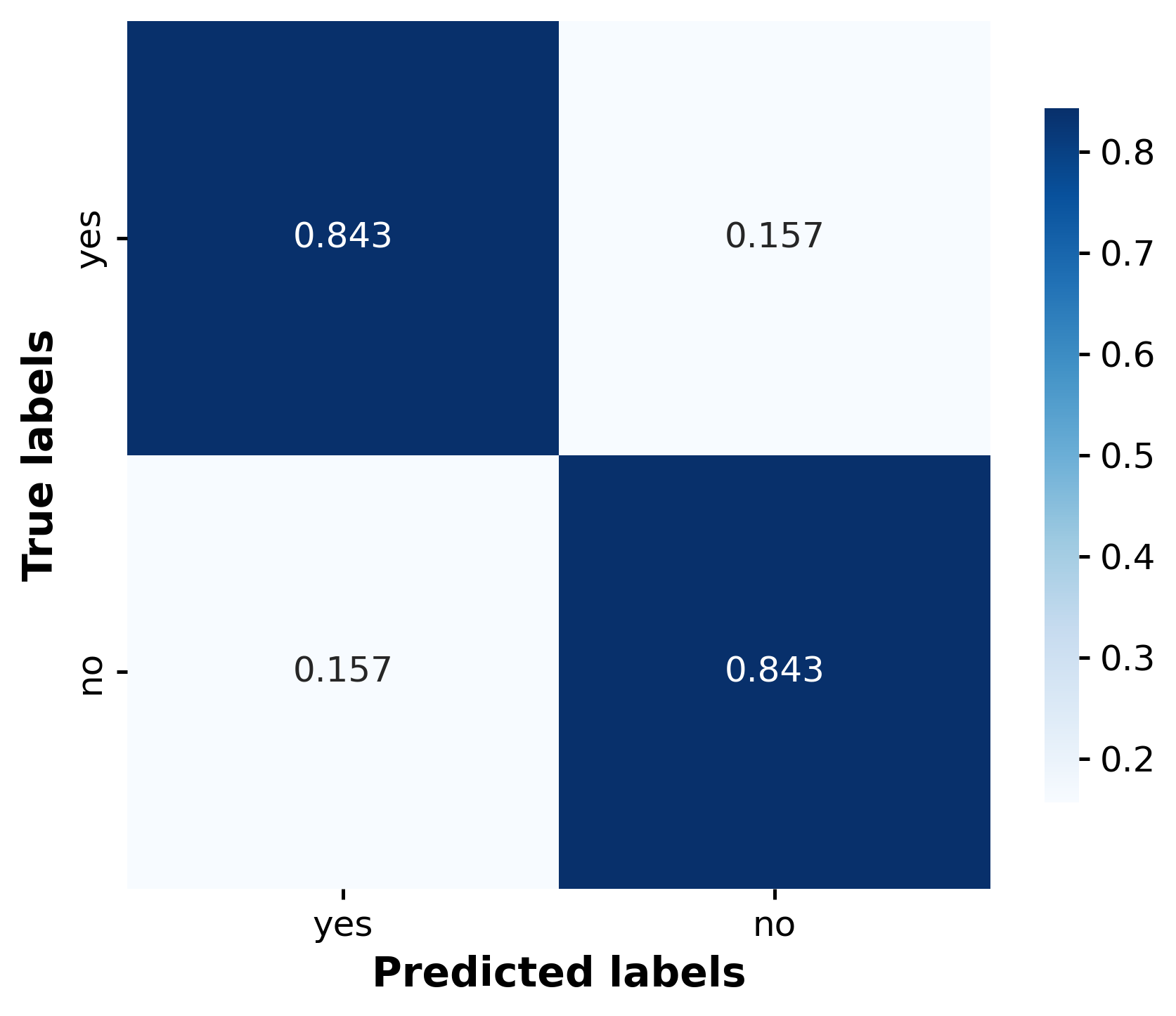}
      \caption{TAD ($\alpha=0.5$)}
    \end{subfigure}\hfill
    \begin{subfigure}[t]{0.15\textwidth}
      \centering
      \includegraphics[width=\linewidth]{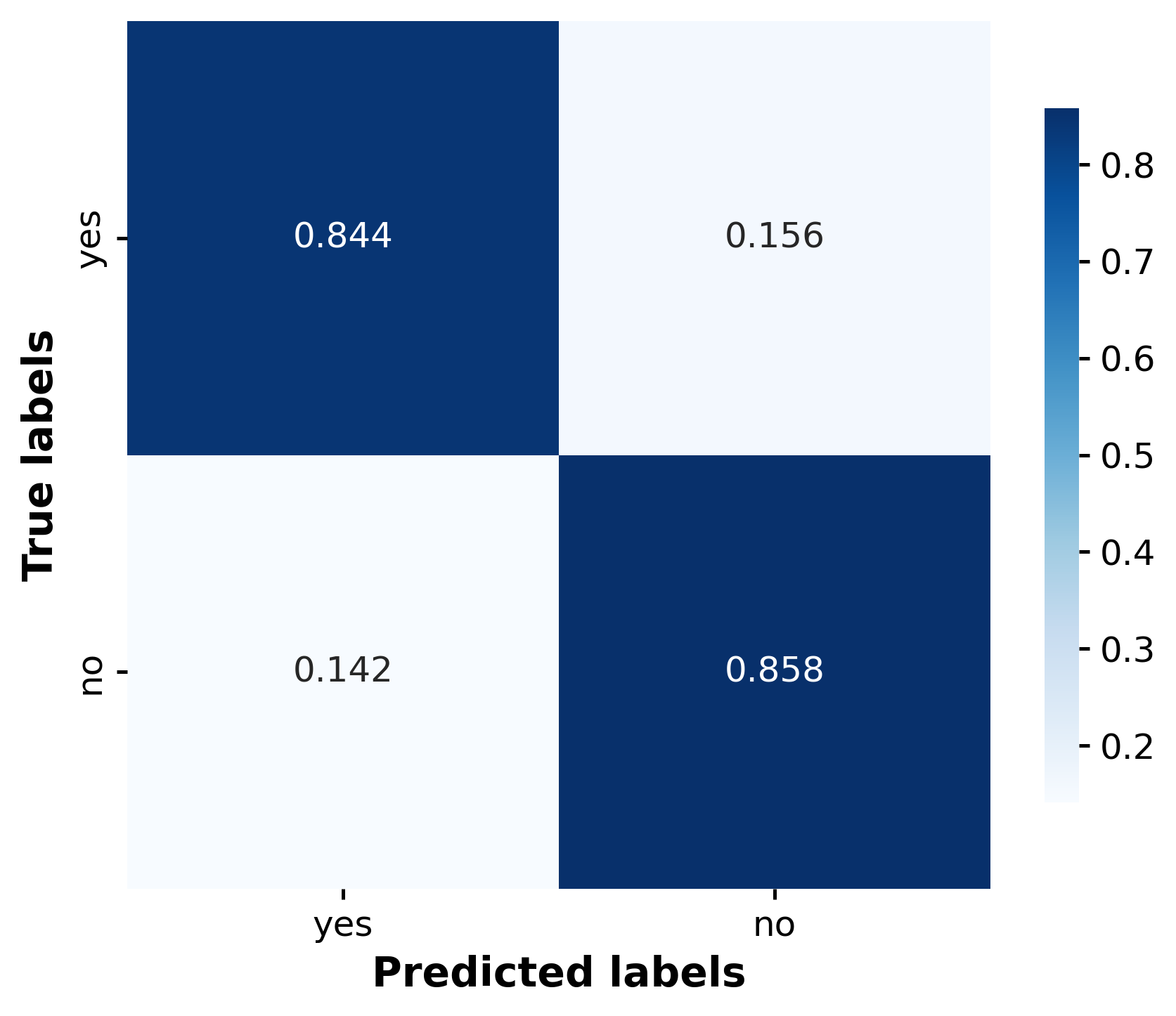}
      \caption{TAD ($\alpha=1.0$)}
    \end{subfigure}
    \hspace*{\fill}

  \caption{Confusion matrices on AudioCaps-Hallucination for Qwen2-Audio-7B-Instruct under the \textsc{Random} setting: Default (a), AAD ((b) and (c)), and TAD ((d) and (e)).}
  \vspace{-1.0em}
  \label{fig:audicaps_random_confmat}
\end{figure}

Figure~\ref{fig:audicaps_random_confmat} compares the normalized confusion matrices of Qwen2-Audio-7B-Instruct on AudioCaps-Hallucination under default decoding, AAD, and TAD. Since the matrices are normalized over true labels, the diagonal entries correspond to class-wise recall. Default decoding exhibits a pronounced affirmative bias, with recall for true-\textsc{Yes} reaching 0.920 while recall for true-\textsc{No} is only 0.266, and most true-\textsc{No} instances misclassified as \textsc{Yes} at 0.734. AAD largely preserves recall for true-\textsc{Yes} at 0.911 while improving recall for true-\textsc{No} to 0.655, consistent with suppressing prior-driven affirmations. TAD further increases recall for true-\textsc{No} to 0.858, corresponding to a hallucination error of 0.142, with a modest decrease in recall for true-\textsc{Yes} to 0.844. This reflects a more conservative and evidence-dependent decision boundary that yields a larger gain in rejecting unsupported objects than the loss on true positives.

\vspace{-0.5em}

\subsubsection{ROC and AUC}
To probe how TAD suppresses hallucination at the decision point, we analyze ROC curves using only the first-step next-token logits. Given $\mathbf{z}\in\mathbb{R}^{|V|}$, we pool evidence for \textsc{Yes}/\textsc{No} over token sets via log-sum-exp:
\begin{equation}
L_{\text{yes}}=\log\sum_{i\in S_{\text{yes}}}\exp(z_i),\qquad
L_{\text{no}}=\log\sum_{i\in S_{\text{no}}}\exp(z_i),
\end{equation}
and define a continuous score
\begin{equation}
P(\text{no})=\frac{\exp(L_{\text{no}})}{\exp(L_{\text{no}})+\exp(L_{\text{yes}})},
\end{equation}
which is treated as the decision variable with \textsc{No} as the positive class. By sweeping a threshold on $P(\text{no})$, we obtain ROC curves (TPR vs.\ FPR) that reflect early-stage discriminability independent of later decoding. As shown in Figure 3, TAD achieves higher ROC and AUC than default decoding and AAD, indicating stronger audio-grounded evidence at the first step and reduced reliance on language priors.

\begin{figure}[t]
  \centering
    \hspace*{\fill}
    \begin{subfigure}[t]{0.23\textwidth}
      \centering
      \includegraphics[width=\linewidth]{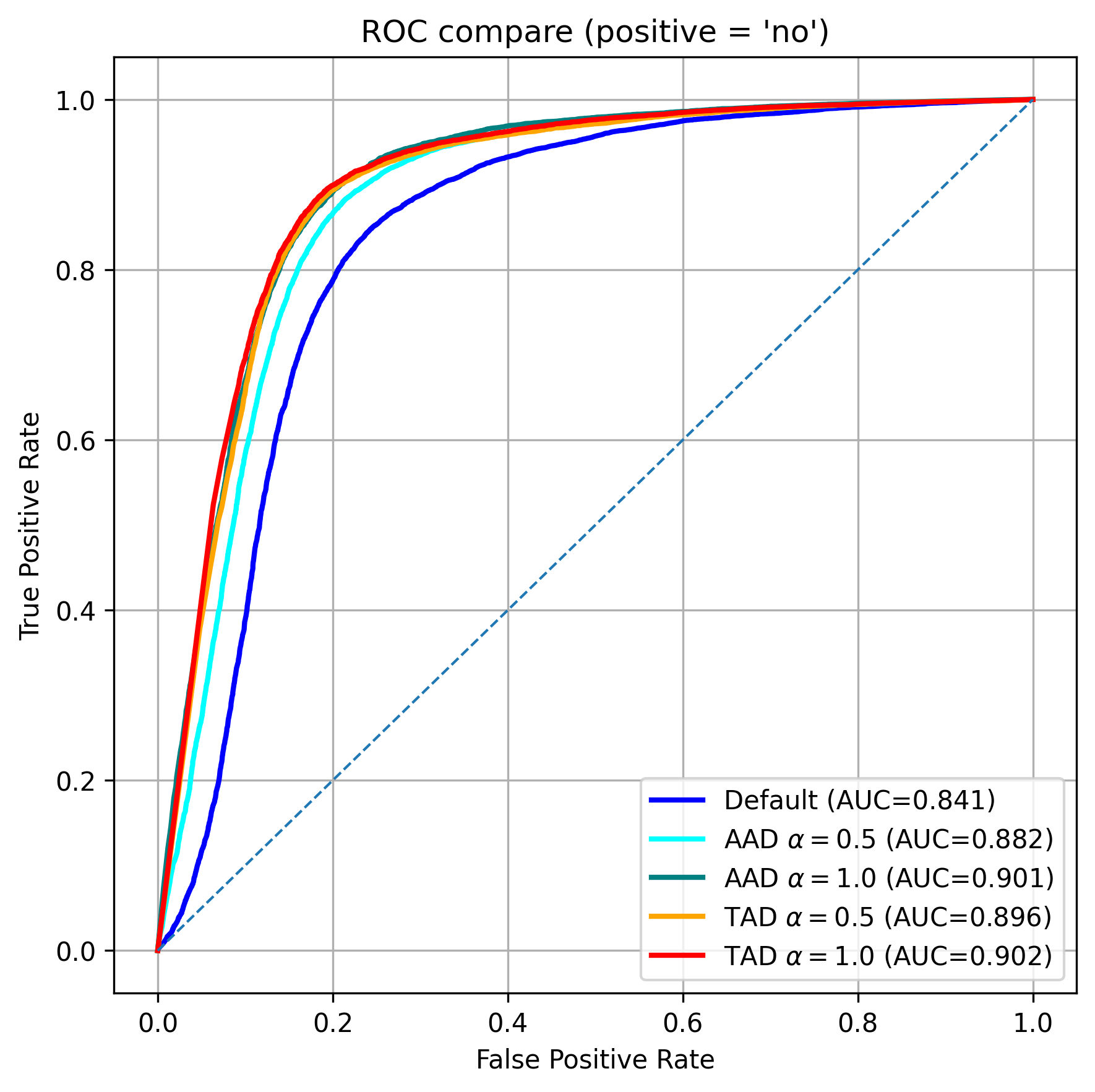}
    \end{subfigure}\hfill
    \begin{subfigure}[t]{0.23\textwidth}
      \centering
      \includegraphics[width=\linewidth]{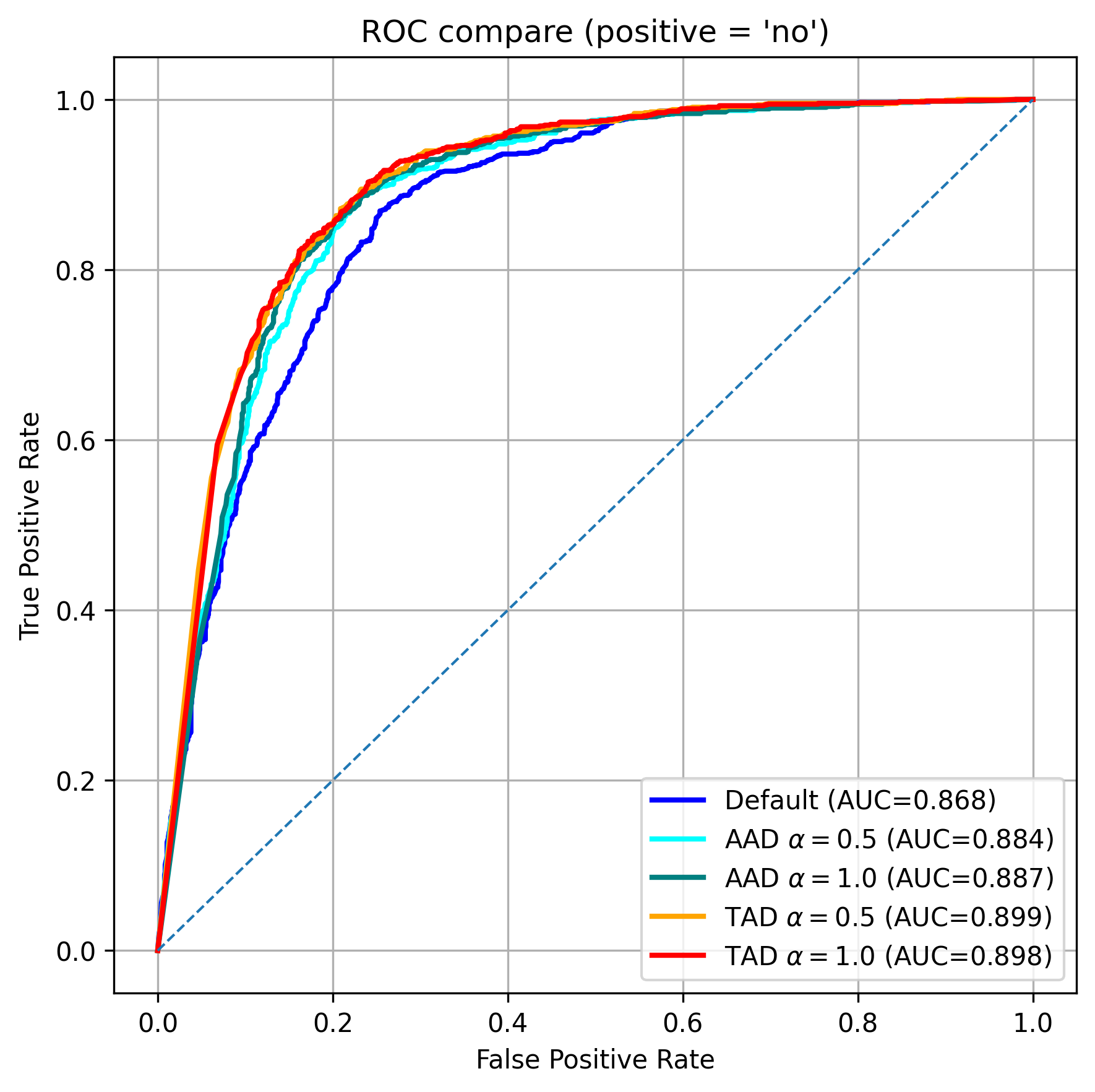}
    \end{subfigure}
    \hspace*{\fill}
  \caption{ROC curves and AUC values of Qwen2-Audio-7B-Instruct on AudioCaps-Hallucination under the \textsc{Random} setting (left) and on Clotho-AQA (right).}
  \vspace{-0.5em}
  \label{fig:roc}
\end{figure}

\subsubsection{Quantitative Behavior of Audio Evidence Scores}
Figure~4 visualizes TAD’s first-step evidence in two complementary views: the histograms in (a) and (c) compare the empirical distribution of $\delta$ for true-\textsc{Yes} versus true-\textsc{No}, while the scatter plots in (b) and (d) contrast each example’s with-audio margin against its silent-reference margin to reveal how audio reshapes the initial decision. In both AudioCaps-Hallucination and Clotho-AQA, the scatter plots show that the silent-reference margin is positive for most samples and largely overlaps between labels, exposing a strong, label-agnostic affirmative prior. By contrast, the with-audio margin separates the classes more clearly, with true-\textsc{Yes} concentrating at higher values and true-\textsc{No} shifting downward. Consistently, the $\delta$ histograms exhibit a boundary near zero, with true-\textsc{No} skewing toward $\delta<0$ and true-\textsc{Yes} toward $\delta>0$, supporting a small gating threshold and explaining TAD’s behavior: it preserves affirmative answers when audio provides a clear margin gain, but selectively suppresses unsupported \textsc{Yes} tendencies when the gain is weak, thereby reducing hallucinated affirmations.

\begin{figure}[t]
  \centering

  \begin{subfigure}[t]{0.48\linewidth}
    \centering
    \includegraphics[width=\linewidth]{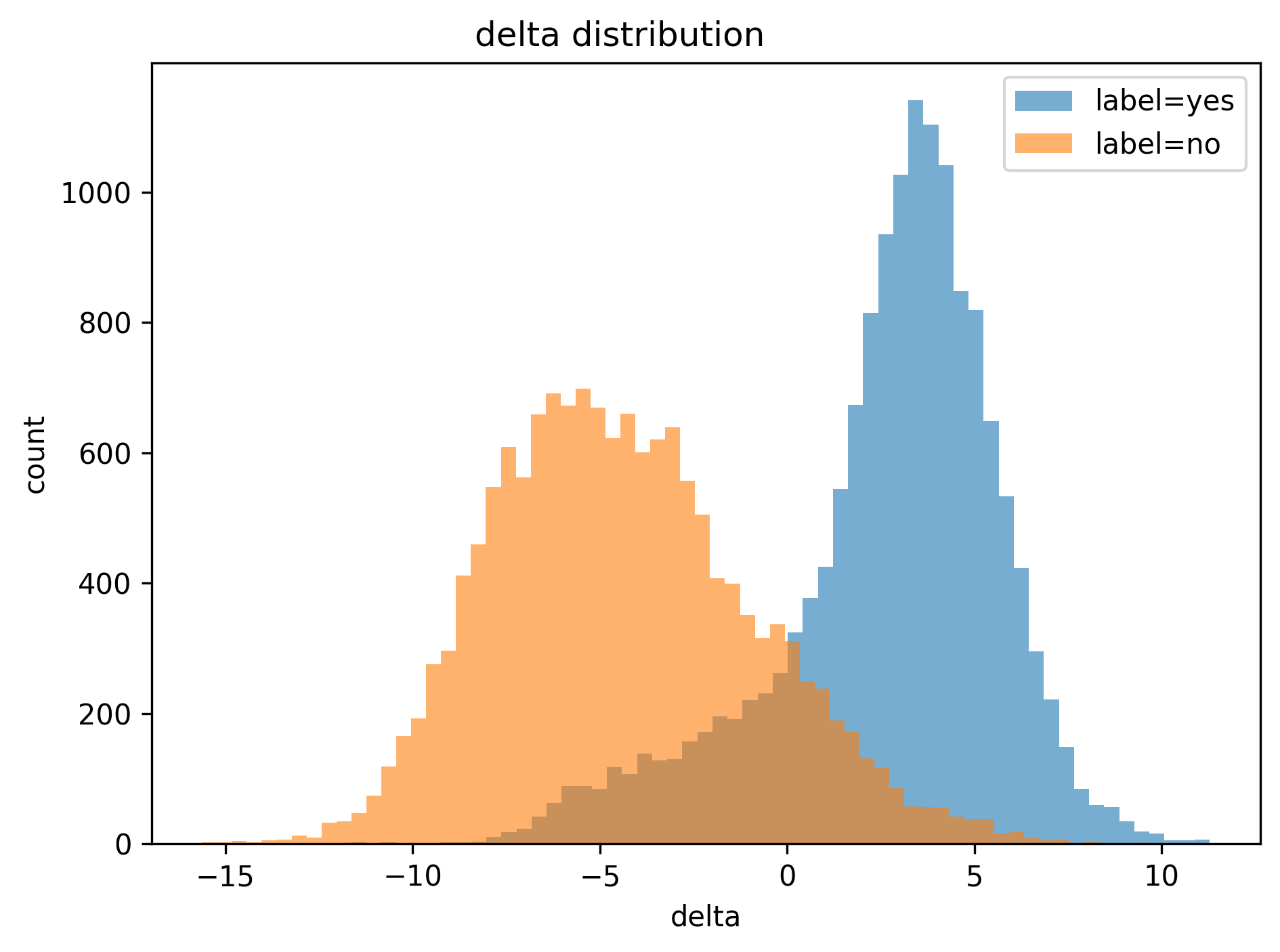}
    \caption{}
    \label{fig:fig4_1}
  \end{subfigure}\hfill
  \begin{subfigure}[t]{0.48\linewidth}
    \centering
    \includegraphics[width=\linewidth]{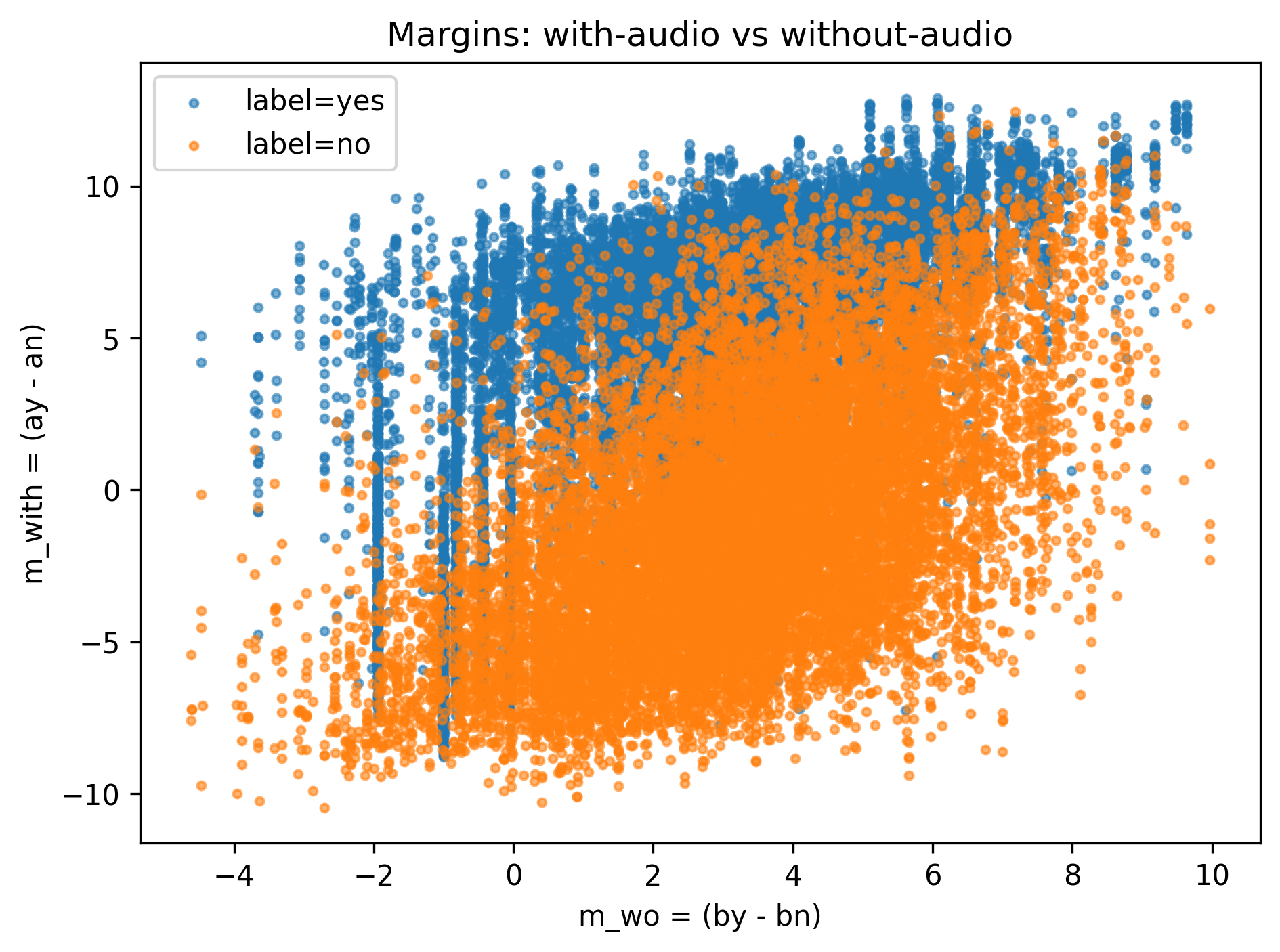}
    \caption{}
    \label{fig:fig4_2}
  \end{subfigure}

  \vspace{0.1em}

  \begin{subfigure}[t]{0.48\linewidth}
    \centering
    \includegraphics[width=\linewidth]{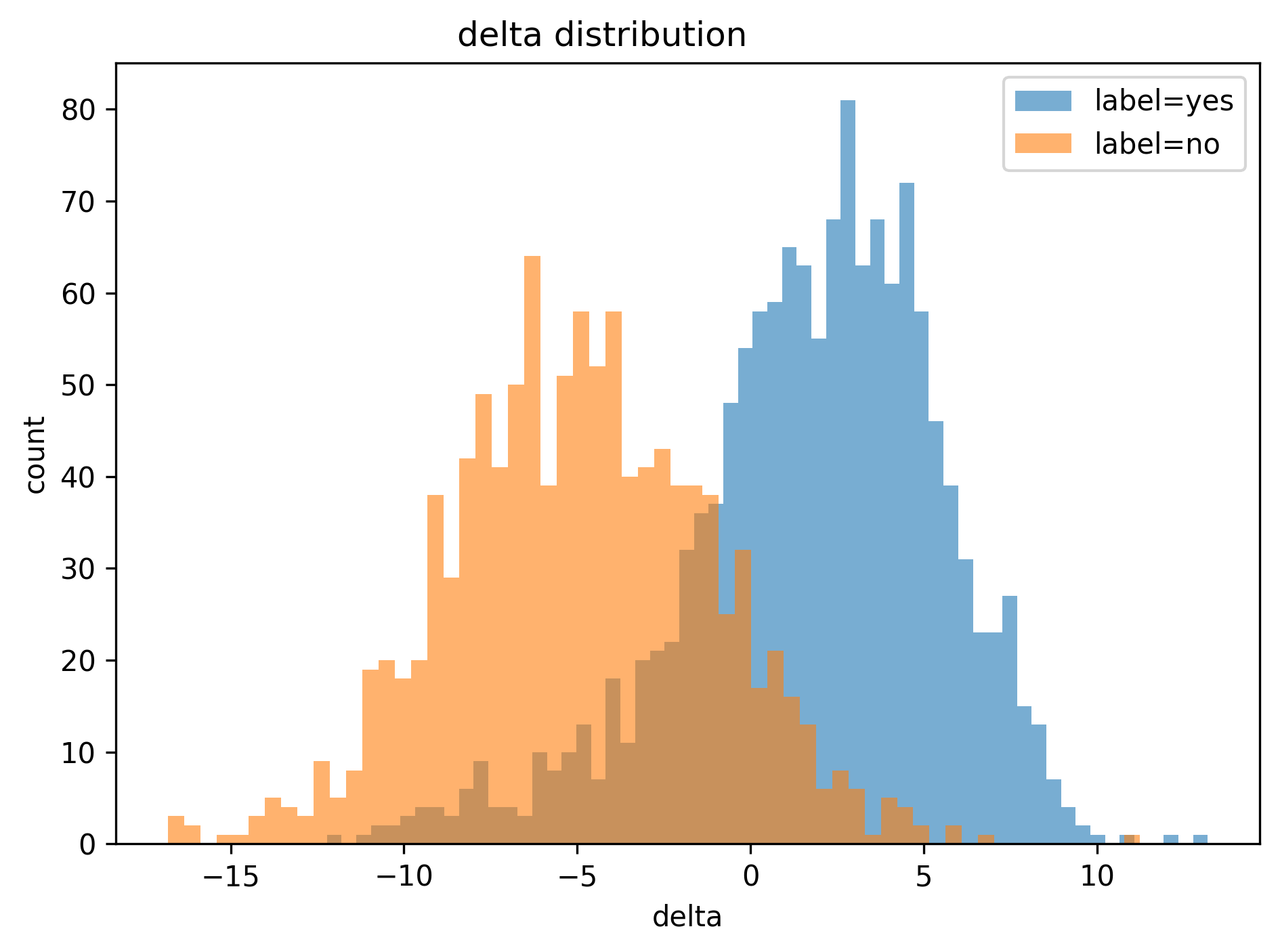}
    \caption{}
    \label{fig:fig4_3}
  \end{subfigure}\hfill
  \begin{subfigure}[t]{0.48\linewidth}
    \centering
    \includegraphics[width=\linewidth]{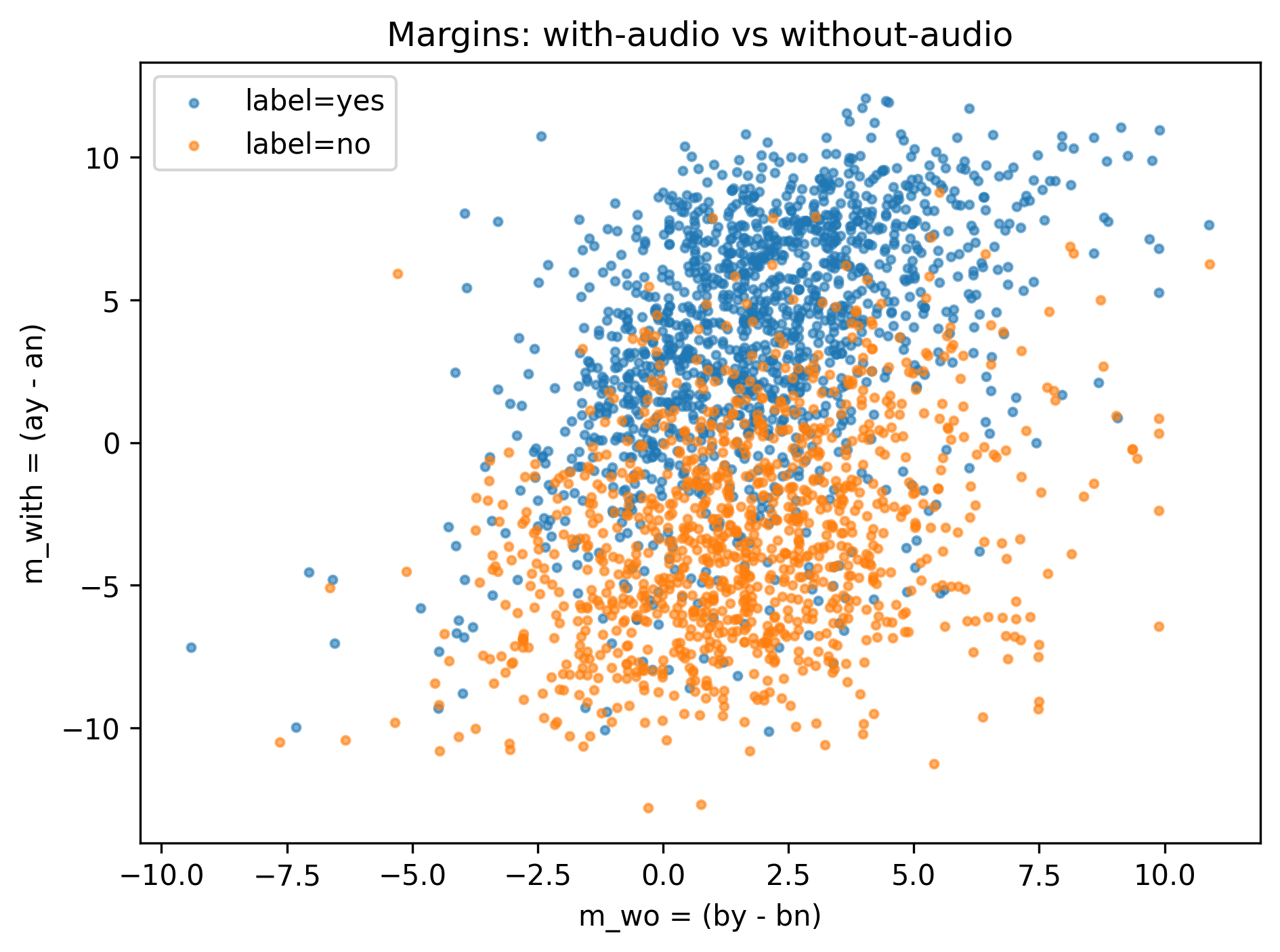}
    \caption{}
    \label{fig:fig4_4}
  \end{subfigure}

  \caption{Visual Analysis of Audio Evidence on AudioCaps-Hallucination ((a) and (b)) and Clotho-AQA ((c) and (d)).}
  \vspace{-1.5em}
  \label{fig:fig4_grid}
\end{figure}

\section{Conclusion}


We propose TAD, a training-free, decision-aware method to mitigate audio object hallucination in LALMs. TAD estimates audio-supported evidence and applies a confidence-gated adjustment at the first decoding step to downweight unsupported affirmative outputs, preserving stable responses while reducing prior-driven yes bias in binary AQA. Experiments on two datasets show improved robustness over prior methods, and analyses of early-step evidence and separability help explain the gains from this targeted, class-conditional intervention. 

\section{Acknowledgments}
This work was supported in part by the Henan Province Major Industrial ``Challenge-Based Innovation'' under Grant 251000210300, in part by the Natural Science Foundation of Henan under Grant 252300420990, and in part by the Science and Technology Key Project Plan of Henan under Grant 252102211040.

\section{Generative AI Use Disclosure}
Generative AI tools were used solely to assist with language polishing and readability improvement. The authors take full responsibility for the research design, experimental process, data analysis, result interpretation, and scientific conclusions presented in this manuscript.

\bibliographystyle{IEEEtran}
\bibliography{mybib}

\end{document}